\documentclass[usenatbib,useAMS,a4]{mn2e}
\usepackage{graphicx} \usepackage{natbib}
\usepackage{lastpage} \usepackage{subfigure}
\usepackage{float}
\let\oldhat\hat

\renewcommand{\hat}[1]{\oldhat{\mathbf{#1}}}

\title[On Blue Loops]{On the Blue Loops of Intermediate-Mass Stars}

\author[J. J. Walmswell, C. A. Tout and J. J. Eldridge]{J. J. Walmswell$^1$ \thanks{E-mail: jwalmswell@gmail.com}, C. A. Tout$^1$ and J. J. Eldridge$^2$\\
$^1$Institute of Astronomy, Madingley Rd, Cambridge, CB3 0HA, UK\\
$^2$Department of Physics, The University of Auckland, Private Bag 92019, Auckland, New Zealand}

\date{February 2015} \pagerange{\pageref{firstpage}--\pageref{lastpage}}
\begin{document}

\maketitle
\label{firstpage}

\begin{abstract}
  We consider the blue loops in the Hertzsprung-Russell diagram that
  occur when intermediate-mass stars begin core helium burning. It has
  long been known that the excess of helium above the burning shell,
  the result of the contraction of the convective core during core
  hydrogen burning, has the effect of making such stars redder and
  larger than they would be otherwise. The outward motion of the burning
  shell in mass removes this excess and triggers the loop. Hitherto
  nobody has attempted to demonstrate why the excess helium has this
  effect. We consider the effect of the local opacity, which is
  reduced by excess helium, the shell fuel supply, which is also
  reduced, and the local mean molecular weight, which is increased. We
  demonstrate that the mean molecular weight is the decisive reddening
  factor. The opacity has a much smaller effect and a reduced fuel
  supply actually favours blueward motion.
\end{abstract}


\section{Introduction}

Blue loops occur when stars with masses between about $3$~and
$9\,M_\odot$ start to burn helium in their cores. At this point the
star has usually developed a convective envelope and is rising up the
Hayashi line in the Hertzsprung-Russel (HR) diagram. It consists of a
dense core and a diffuse envelope and already possesses an energy
source in the form of a thin hydrogen-burning shell just outside the
core. The envelope is generally not homogeneous and has an excess of
helium in the region above the burning shell. This is the result of
the shrinking extent of the convective core during the earlier core
hydrogen-burning phase. The core comprises a convective inner core, in
which helium is gradually converted to carbon and oxygen, and a thick
shell of radiative helium. The existence of two sources of fusion
energy means that the star has two nuclear time-scales. How these
compare will determine the evolution of the star during this phase.

\begin{figure}
\centering
\includegraphics[width=0.5\textwidth]{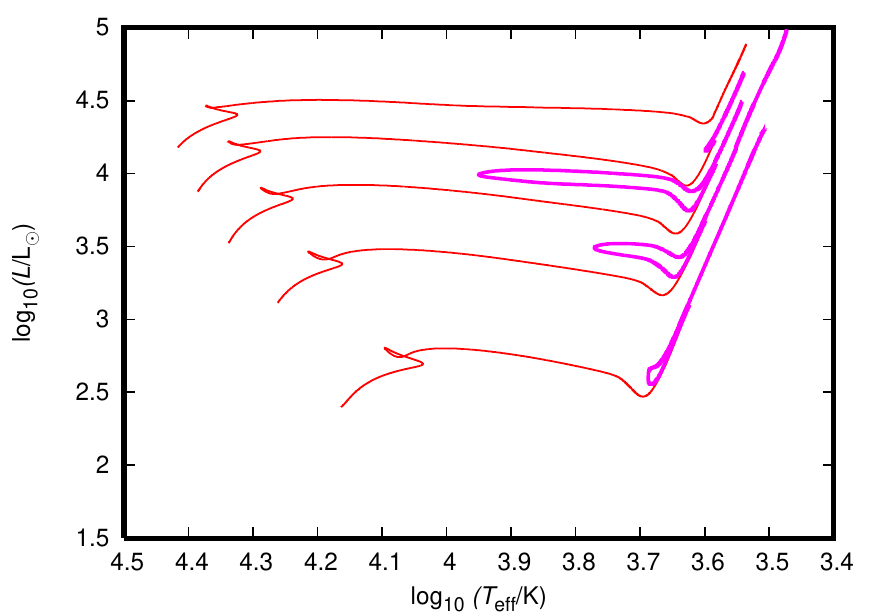}
\caption{Evolution tracks for stars with masses of $4$, $6$, $8$, $10$
  and $12\,M_\odot$. Composition is solar and the blue loops and
  subsequent evolution are indicated by thicker lines. The $12$
  $M_\odot$ star does not at any point move back down the Hayashi line
  and is not considered to have looped.}
\label{fig:Blueloop}
\end{figure}

Figure ~\ref{fig:Blueloop} shows the evolution tracks for five stars with
the traditional solar composition, that is, the hydrogen mass fraction
$X$ is $0.70$, the helium mass fraction $Y$ is $0.28$ and the
metallicity $Z$ is $0.02$. It should be noted that the current best
estimate of solar metalicity is $0.0126$ \citep{2004A&A...417..751A},
although there is still much uncertainty
\citep[e.g.][]{2014arXiv1409.6910S}. All evolution after the tip of
the first ascent of the Hayashi line is rendered with thicker lines to
emphasise the blue loops. Those stars that perform blue loops contract
and heat up, moving leftward in the HR diagram. Eventually they reach
a maximum effective temperature and return to the Hayashi line. The
considerable variation in luminosity is largely due to changes in the
output of the hydrogen-burning shell. The helium luminosity is small
in comparison and merely increases monotonically
\citep{2004A&A...418..213X}. In a nutshell a blue loop consists of a
period of leftward motion in the HR diagram followed by a period of
rightward motion. Note that the most massive star to fully loop is an
$8\,M_\odot$ star. The $10\,M_\odot$ star moves slightly down the
Hayashi line in a vestigial loop and the $12\,M_\odot$ star does not
loop at all. This general behaviour, though it occurs over
slightly-different mass ranges, is observable across the current set
of stellar evolutionary codes and reasonable input physics.

\begin{figure}
\centering
\includegraphics[width=0.5\textwidth]{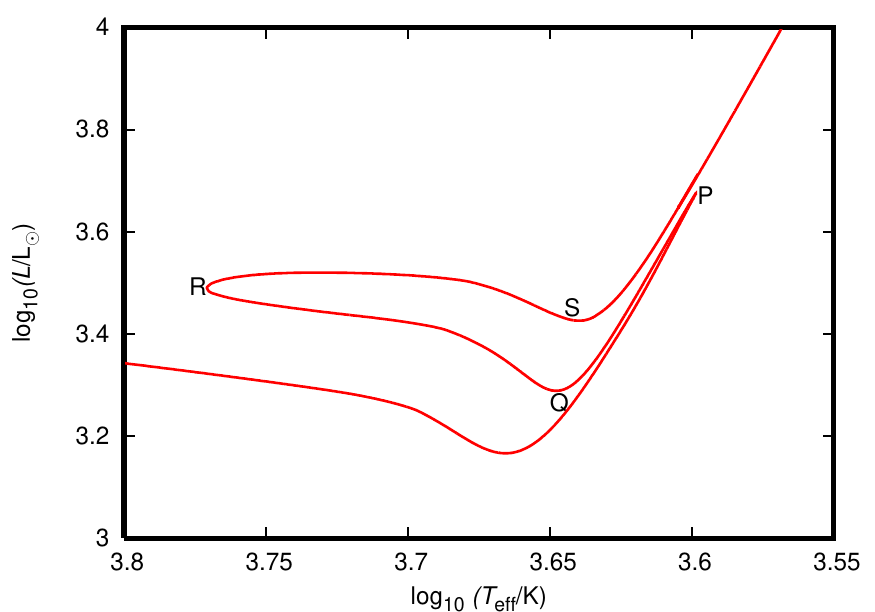}
\caption{The blue loop of a $6\,M_\odot$ star with solar
  composition. Four notable points are labelled: P is the tip of the
  first ascent of the Hayashi line, Q is the subsequent luminosity
  minimum, R is the temperature maximum and S is the second luminosity
  minimum. S marks the end of the loop.}
\label{fig:Blueloopdetail}
\end{figure}

\begin{figure}
\centering
\includegraphics[width=0.45\textwidth]{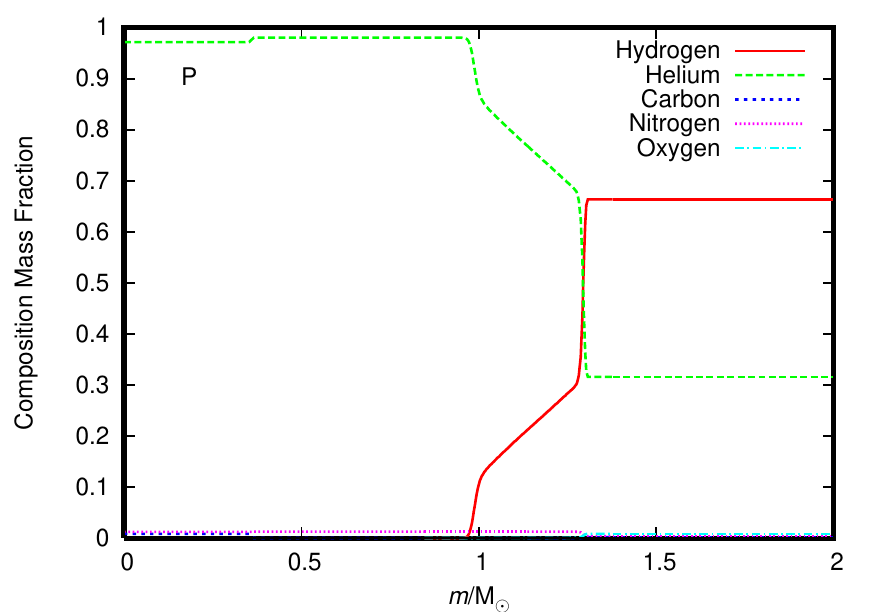}
\includegraphics[width=0.45\textwidth]{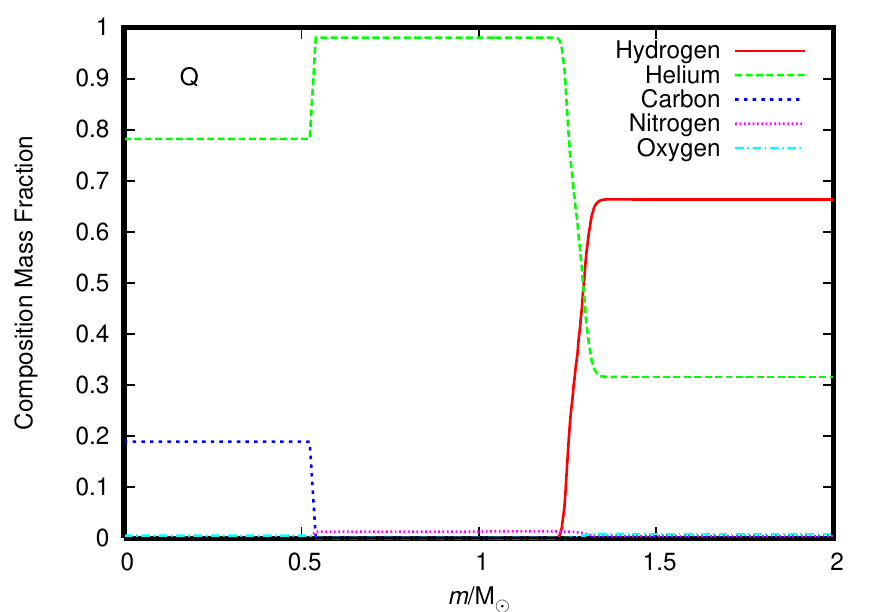}
\includegraphics[width=0.45\textwidth]{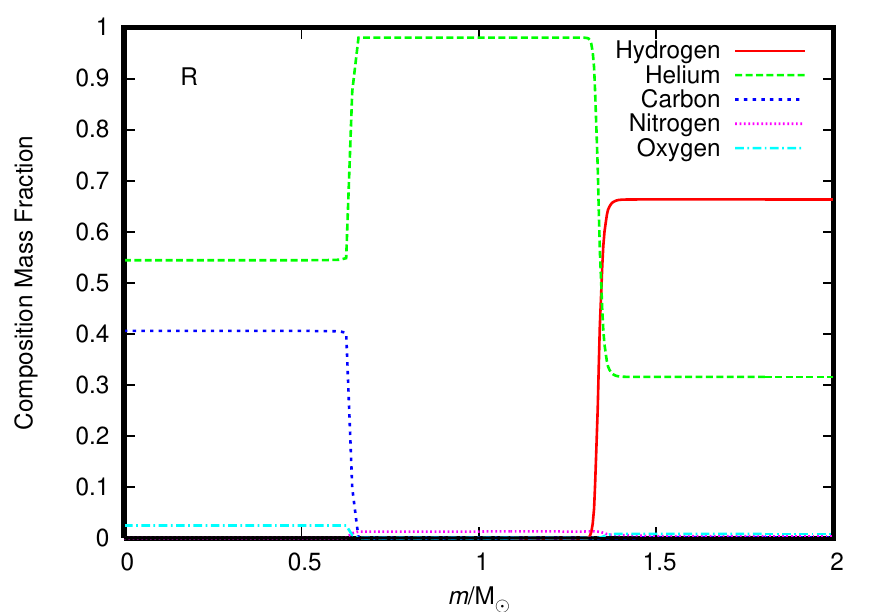}
\includegraphics[width=0.45\textwidth]{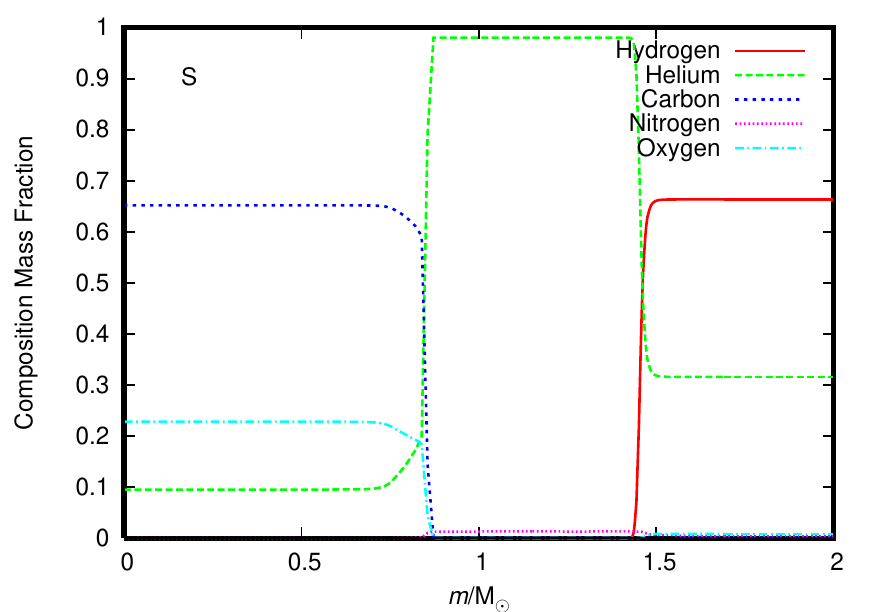}
\caption{Composition profiles of a $6\,M_\odot$ star at the points
  labelled in Fig.~\ref{fig:Blueloopdetail}. P occurs at $67$ Myr, Q at
  $69$ Myr, R at $71$ Myr and S at $76$ Myr.}
\label{fig:Blueloopdetailcompos}
\end{figure}

Figure ~\ref{fig:Blueloopdetail} shows the blue loop of a $6\,M_\odot$
star in greater detail. Four points are labelled as P, Q, R and S and
their corresponding internal compositions are shown in
Fig.~\ref{fig:Blueloopdetailcompos}. In P we can, working outwards,
identify the following features. First, a convective inner core in
which helium is being converted to carbon, secondly the radiative
outer core, thirdly the steep increase in hydrogen that marks the
hydrogen-burning shell, fourthly the steady increase in hydrogen that
marks the retreat of the old convective core and finally the jump to
the envelope composition. This jump is the result of first dredge-up,
that is, the inward penetration of the envelope convective zone into
the region affected by core hydrogen-burning. It truncates the old
composition profile and redistributes the affected matter within the
envelope. In Q we can see that both the inner core and the burning
shell have burnt outwards in mass and this continues in R and S. The
inner core changes to a mixture of helium and carbon and then to a
mixture of carbon and oxygen. The burning shell eventually consumes
the zone of changing composition and reaches the jump. These changes
have consequences that will be discussed in the next section.

In the first part of this paper we examine previous research into the
nature of blue loops. We correct several misconceptions and find that
the removal of the excess helium above the core is the fundamental
explanation for looping. The excess helium alters the stellar
properties in different ways and we investigate them separately. We
conclude by considering how the lengths of blue loops vary with the
initial helium abundance and the metallicity. We find that the
situation is more complicated than is generally assumed.

\section{Previous Work}

The first thorough studies of the blue loop phenomenon were by
\citet*{1971A&A....10...97L} and \citet{1972MNRAS.156..129F}. Both
groups began with static shell-burning models with so-called step
profiles, models for which the composition of the envelope is uniform
from the surface to the core. \citet{1971A&A....10...97L} treated the
core as entirely arbitrary by using $R=R_{\rm Core}$,
$M=M_{\rm Core}$ and $L=L_{\rm Core}$ as inner boundary
conditions. They found that an increase in core mass at fixed core
radius causes rightward motion across the Hertzsprung Gap and up the
Hayashi line. An increase in core radius at fixed core mass has the
opposite effect. The position of a model in the HR diagram was
therefore characterised by the core potential $\Phi=M_{\rm
  Core}/R_{\rm Core}$.  \citet{1972MNRAS.156..129F} considered static
models of complete stars but came to an equivalent conclusion with
regards to their position function $\Lambda^{*}=R_{\rm Core}/M_{\rm
  Core}$.

Evolution alters the core in two ways. First there is the gain of
material from hydrogen burning, secondly there are changes in
composition due to helium burning. \citet{1972MNRAS.156..129F}
considered how $R_{\rm Core}$ varies as a function of composition at
fixed core mass and found a maximum at about $Y=0.35$. This means that
the chemical evolution of the core initially favours an expansion in
core radius and hence leftward motion in the HR diagram. The behaviour
reverses when $Y$ falls below $0.35$ and chemical evolution then
favours core contraction, which becomes rapid when
$Y<0.1$. \citet{1972MNRAS.156..129F} presented a theoretical
justification based on the virial theorem in an appendix.

The dependence of $R_{\rm Core}$ on composition is the correct
behaviour to begin and end a blue loop. However, as
\citet{1971A&A....10...97L}, point out, the magnitude of these changes
to core radius is not nearly enough to account for observed loops. The
overall core mass also increases monotonically during helium burning
and this suppresses looping behaviour entirely. It is therefore
apparent that the expected changes to the core cannot induce
substantial loops, though they eventually initiate the redward motion
that curtails them.

Both \citet{1971A&A....10...97L} and \citet{1972MNRAS.156..129F} came
to the conclusion that the outward growth in mass of the
hydrogen-burning shell is the cause of looping. This outward motion
takes the shell through the composition profile left behind by the
retreat of the convective core, as a comparison of Cases P and Q in
Fig.~\ref{fig:Blueloopdetailcompos} demonstrates. It thus causes the
excess helium above the shell to decline monotonically. By contrast,
both groups of authors found that adding excess helium to the region
above the shell in the step profile models caused them to move
rightwards in the HR diagram: the blue loop in reverse. The effect was
most pronounced when the excess helium was closest to the core but it
persisted beyond the extent of the burning
shell. \citet{1972ApJ...173..631R} agreed and demonstrated that his
models reached their bluest extent in the HR diagram at the point when
the composition at the burning shell finally equalled that of the
stellar surface. Whether or not a star actually performs a blue loop
depends on whether this effect manages to temporarily overcome the
consequences of the core evolution. It should be noted that, as can be
seen in Fig.~\ref{fig:Blueloopdetailcompos}, the composition change
between the envelope and the core never becomes entirely
discontinuous. This is not permitted by the star and a
perfectly vertical step profile model always relaxes somewhat to give
something like Case S.

As we have already observed, the evolutionary composition profile is
complicated by the effects of first dredge-up. This removes the outer
part of the profile and replaces it with a near-vertical jump that
marks the innermost point of convective penetration. The excess helium
above the burning shell is thus reduced (see Case P in
Fig.~\ref{fig:Blueloopdetailcompos}) and therefore anything that moves
this discontinuity closer to the core produces a more extensive
loop. \citet{1972ApJ...173..631R} showed this for his increased
envelope opacities, which had the effect of promoting deeper
convection zones.  \citet{1991ApJ...374..288S} showed this for
envelope convective overshooting. Stars more massive than about
$9\,M_\odot$ do not experience significant inward penetration of the
convective envelope at solar metallicity and thus looping is limited
to no more than slight motion down and up the Hayashi line.

The fact that excess helium above the burning shell compels a
shell-burning star to be larger and redder than it would otherwise be
is beyond dispute. The reader can verify this with a stellar
evolution code of their choice. Perhaps as a consequence blue loops
have received little attention since the seventies, despite the fact
that the details of this reddening mechanism are unknown. To attempt
to disentangle the chains of cause and effect involved in stellar
evolution is not always straightforward.

\citet{1992ApJ...400..280R} explained the expansion of
intermediate-mass and massive stars to red giants as a consequence of
a runaway thermal instability in the envelope. The expansion of the
envelope causes it to cool, the heavy ions recombine and this
increases the opacity. The envelope absorbs more energy and expands
further on the thermal time-scale. The initial trigger is an increase
in luminosity and similar arguments were used to explain motion in the
reverse direction in the HR diagram via a drop in luminosity and
subsequent contraction. This drop in luminosity was said by
\citet{1990ApJ...364..527S} to be the result of the increase in local
opacity as the ratio of helium to hydrogen above the burning shell
decreases. \citet{1992ApJ...400..280R} claimed that a blue loop would
be triggered if the stellar luminosity fell below a certain threshold.

\begin{figure}
\centering
\includegraphics[width=0.5\textwidth]{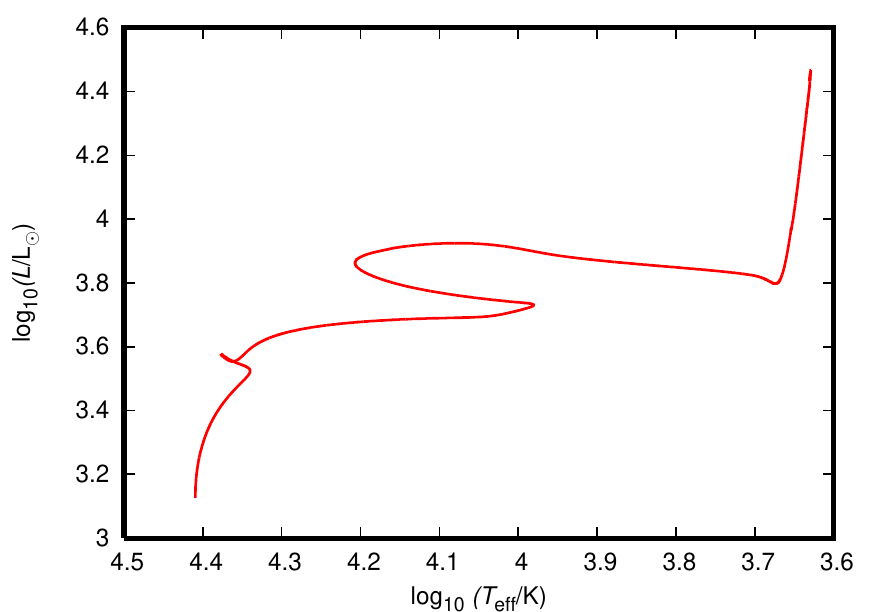}
\caption{Evolution track in the HR diagram of a $6\,M_\odot$ star with $Z=0.0001$.}
\label{fig:Bluelooplowz}
\end{figure}

The overall theory has been heavily criticised
\citep{2005slfh.book..149F}. In the case of the blue loop it can be
readily disproved by turning our attention to stars with very low
metallicities. If sufficiently massive these stars start helium
burning in the Hertzsprung gap and perform a blue loop without either
ascending the Hayashi line or experiencing a drop in luminosity
(Fig.~\ref{fig:Bluelooplowz}). \citet{1992ApJ...400..280R} have
mistaken correlation for causation. If a star on the Hayashi line is
to perform a blue loop then the motion back down the line necessarily
requires a drop in luminosity if the temperature is to
increase. Similarly the thermal-instability explanation confuses cause
and effect. If the interior of a star changes more rapidly than the
envelope can adjust then it must indeed be thermally unstable. This
instability does not itself cause anything. In any case stars between
about $3$ and $5\,M_\odot$ perform blue loops without any thermal
instability whatsoever.

In a similarly-mistaken vein \citet{2004A&A...418..213X} argued that
whether or not a star performs a loop depends on whether the envelope
is dominated by radiation or convection at the post helium-ignition
luminosity minimum. They argued that a convective envelope responds to
increased luminosity by developing more extensive convective zones and
expanding back up the Hayashi line. A radiative envelope instead
improves its conductivity by heating up and moving into the
Hertzsprung gap.  They suggested that a possible criterion is for the
ratio of the convective envelope mass to the total envelope mass to
fall below some minimum at the luminosity minimum. For solar
metallicity they estimated this minimum to be between $0.3$ and
$0.45$.

Again cause and effect are confused. Convection is a response made by
the envelope if it is compelled to expand beyond a certain point by
the actions of the core. In no sense is it an independent property of
the envelope. During both the motion down the Hayashi line and into
the Hertzsprung gap the star shrinks and becomes hotter
monotonically. The whole misconception appears to have arisen from a
misreading of a paragraph written by \citet{1992ARA&A..30..235C}:
\emph{`...a rapid contraction of the envelope readjusts the outer
  layers from convective to radiative and the star moves to the second
  region, where the remaining part of the core He-burning phase
  occurs. This causes the blue loops.'} This is true in the precise
sense that a star cannot execute a substantial loop, i.e. greatly
increase its effective temperature, without leaving the Hayashi
line. This departure must necessarily be associated with the end of
convection in the envelope.

Little more needs to be said about the convective envelope mass
criterion.  Subsequently \citep{2004A&A...418..225X} the authors
found, unsurprisingly, that the criterion depended on stellar mass,
metallicity and helium abundance. It does not point to any deeper
truth.

\citet{2004A&A...418..213X} also made the mistake of attempting to
find discrete explanations for the luminosity minimum and subsequent
increase.  They explained the initial drop in luminosity with what
they called the push effect, the composition-induced initial
expansion of the core. They claimed that this compels the shell to
expand and cool and therefore causes the hydrogen-burning luminosity
to decrease. The subsequent increase in luminosity was explained by
the increased predominance of what they called T prompt and X prompt,
which are respectively the heating of the shell by the
increasingly-hot core and the increased availability of fuel. These
explanations are superficially attractive but are based on little more
than assertion.  In the next section we demonstrate that loops are
actually hindered by the increased fuel supply and do not require any
changes to the core.

Stellar evolution is too complicated to be reduced to glib
explanations. A better approach was that of
\citet{1971A&A....10...97L}, who classified their shell-burning models
into three regimes according the the relative mass of the
envelope. Regime I consists of blue stars with a small envelope mass,
regime II of red stars with an intermediate envelope mass and regime
III of blue stars with a large envelope mass. Regime I contains those
stars that have lost much of their envelope and, though applicable to
Wolf-Rayet stars, is not of relevance to a consideration of blue
loops. Regime III contains blue loop stars that are burning helium in
the Hertzsprung Gap and regime II those stars after they have moved
back to the Hayashi line. \citet{1971A&A....10...97L} found that
altering either the core potential or the excess helium could shift a
star from one regime to another through a series of static models. It
is apparent that excess helium has the effect of making a star that
would be in regime III with a step profile behave like a star in
regime II. The outward motion of the burning shell lessens this effect
and causes the loop.

The different regimes represent different sets of solutions to the
stellar evolution equations for shell-burning stars. If a star is in
regime II then a reduction in luminosity is accompanied by an increase
in temperature as it moves down the Hayashi line and contracts. If it
is in regime~III then an increase in temperature may well be
accompanied by an increase in luminosity. Movement from regime~III to
regime II always involves the expansion and cooling of a star and is
accompanied by a drop in luminosity at the bottom of the Hayashi
line. All that can be said is that there is transition between the
regimes and that some factors induce motion in one direction and some
in the other. In the next section we consider the different
consequences of excess helium.

\section{The Varied Consequences of Excess Helium}

We have established that the only solid fact about blue loops is that
they are the result of the outward motion of the burning shell through
a changing composition profile. This in turn is a consequence of the
fact that excess helium above the burning shell has the effect of
shifting a star to the right in the HR diagram. Removing it, provided
that this happens faster than the evolution of the core, triggers a
blue loop. The presence of the excess helium alters the local stellar
properties in various ways and some of these alterations have been
suggested, although not demonstrated, to be of importance.

The opacity, as invoked by \citet{1992ApJ...400..280R}, has already
been mentioned. The replacement of a mass of helium with an equal mass
of hydrogen results in an increase in the local opacity. This is to be
expected: two electrons and an alpha particle provide less opacity
than four protons and four electrons. It is well known that a
\emph{reduction} in envelope opacity makes stars bluer but this is the
consequence of an alteration to the entire envelope rather than to the
area immediately at and above the burning
shell. \citet{1977ApJ...212..507S} found that if he increased the
thickness of the excess helium region beyond a certain point then his
stars started becoming bluer rather than
redder.

 \citet{2004A&A...418..213X} observed that the replacement of
helium with hydrogen provides more fuel for the burning shell and
suggested that this should increase the luminosity. Finally
\citet{1971A&A....10...97L} pointed to the paper on shell homology by
\citet{1970A&A.....6..426R}, who found that the luminosity of stars on
the Hayashi line depends on a high power of the mean molecular
weight, which is lowered by the replacement of helium with hydrogen.

To test these possibilities we need to make some stellar models. All
the models mentioned in this paper were computed with the Cambridge
${\sc \rm STARS}$ code, which was originally developed by Peter
Eggleton in the 1960s \citep{Eggleton}. It uses a non--Lagrangian
mesh, where a mesh function ensures that the points are distributed so
that between them quantities of physical interest do not vary by a
large amount. The code has been gradually improved and updated and the
version used in this research is based on that described by
\citet{Stancliffe} and references therein.  All our models use 499
mesh points because this was found to be the minimum required to
consistently reproduce the evolutionary phases of interest. Models
with more points are effectively identical but take longer to evolve,
whereas models with fewer points have slightly different evolutionary
paths. Mass loss, a very small effect in this mass range, was turned
off and is ignored throughout this paper unless stated
otherwise. Convective overshooting is obtained with the method of
\citet{Schroder}, with an overshooting parameter of ${\rm
  \delta_{OV}=0.12}$. Overshooting is implemented in this manner
throughout the star, with the consequence that inward overshooting
from the envelope occurs as well as outward overshooting from the core.

There are therefore three candidates: the opacity, the fuel supply or
the mean molecular weight. To test in isolation the effects of these
factors we modified the ${\sc \rm STARS}$ code to include an
additional element which we have called arbitrarium. This was inspired
by the work of \citet{2009PASA...26..203S} who modified their code
such that helium had an atomic mass of 1.5 and thus contributed in the
same way as hydrogen to the mean molecular weight, with the aim of
exploring the role of this quantity in the formation of red
giants. The properties of arbitrarium are, needless to say, arbitrary.

\begin{figure}
\centering
\includegraphics[width=0.45\textwidth]{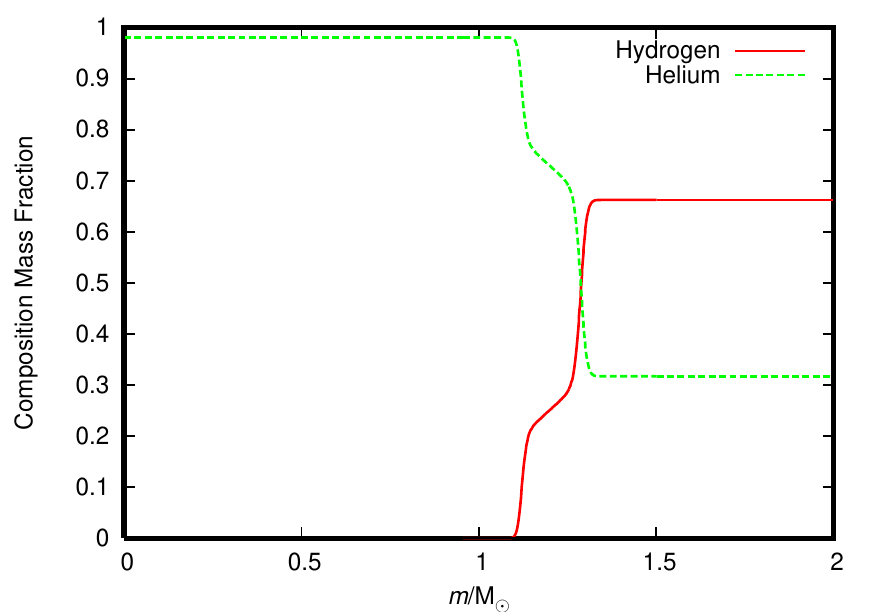}
\caption{Chemical profile for the $6\,M_{\odot}$ initial model.}
\label{fig:prof1}
\includegraphics[width=0.45\textwidth]{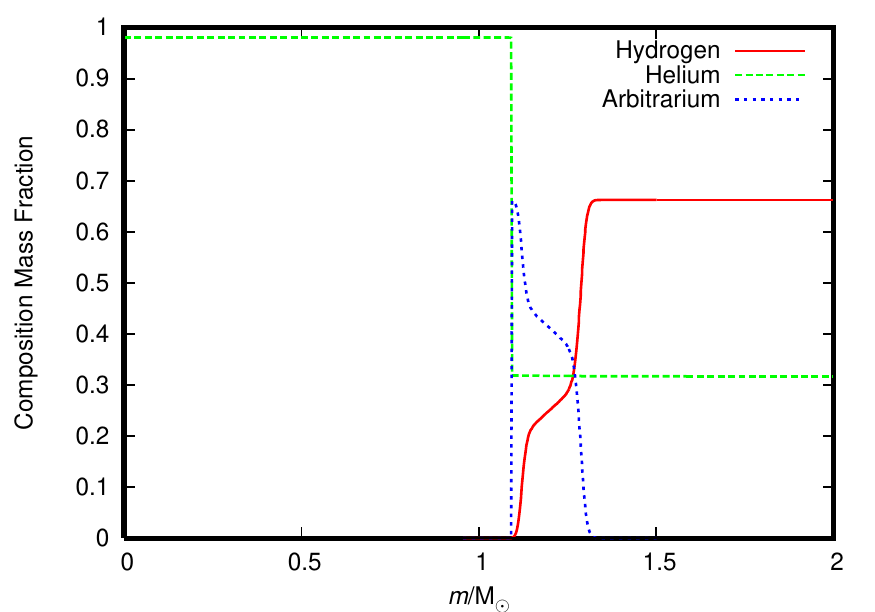}
\caption{Chemical profile for the $6\,M_{\odot}$ initial arbitrarium model.}
\label{fig:prof2}
\includegraphics[width=0.45\textwidth]{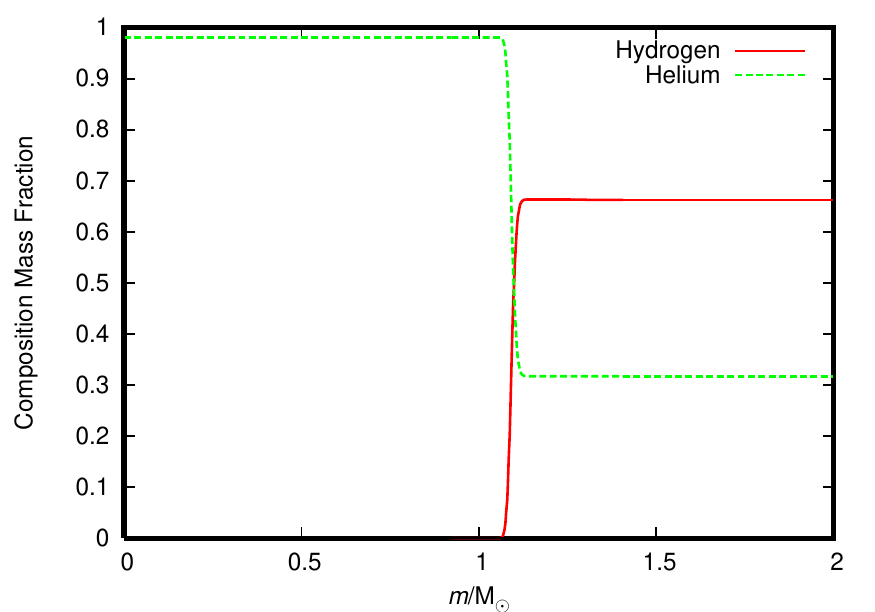}
\caption{Chemical profile for the $6\,M_{\odot}$ step profile model.}
\label{fig:prof3}
\end{figure}

We began with a $6\,M_{\odot}$ star of solar composition and allowed
it to evolve normally until the start of helium burning. At this point
the star consisted of a helium core, a homogeneous envelope and a
transition profile between the two (Fig.~\ref{fig:prof1}). There is an
inner edge at $1.07\,M_{\odot}$ and an outer edge at $1.32$
$M_{\odot}$. The latter is the result of the inward motion of the
convective envelope during first dredge-up and the former occurs
because shell burning is in a very narrow mass range. The intermediate
slope was left behind by the contraction of the convective core during
hydrogen burning. To ensure a pure helium core we froze the chemical
conversion of helium to carbon during the initial model run whilst
leaving the energy generation rates unchanged. The model was extracted
a short but finite time after the start of helium burning, with the
consequence that the helium core is slightly more massive than it was
at the tip of the first ascent of the Hayashi line.

We then replaced the excess helium above the core with arbitrarium
(Fig.~\ref{fig:prof2}) to create an initial model. The arbitrarium was
given the all the properties of helium, so this change had no effect
on the star. For the purpose of comparison we made another model in
which the excess helium was replaced with hydrogen to give a step
profile (Fig.~\ref{fig:prof3}).

\subsection{Opacity}

\begin{figure}
\centering
\includegraphics[width=0.5\textwidth]{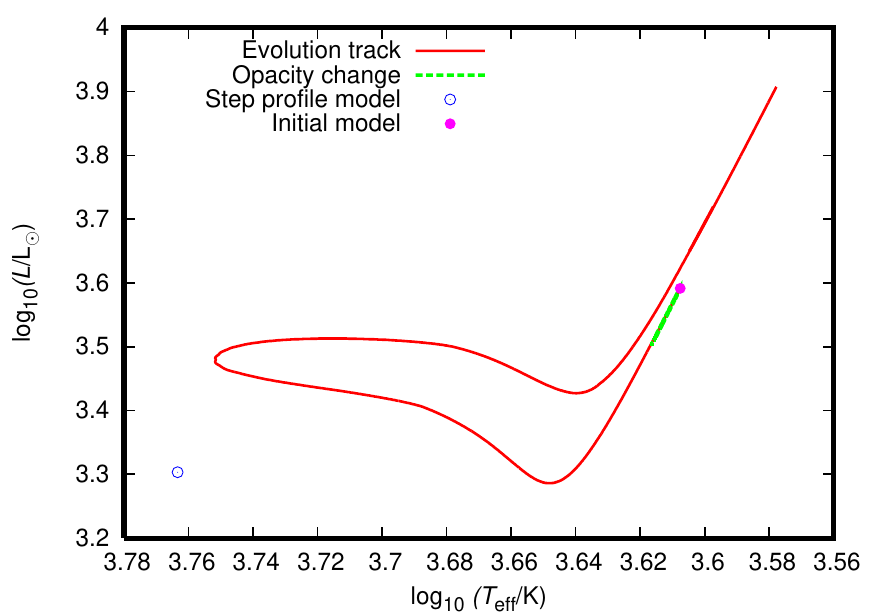}
\caption{The effect of giving the arbitrarium the opacity of
  hydrogen. It is initially identical to helium.}
\label{fig:opaq}
\includegraphics[width=0.5\textwidth]{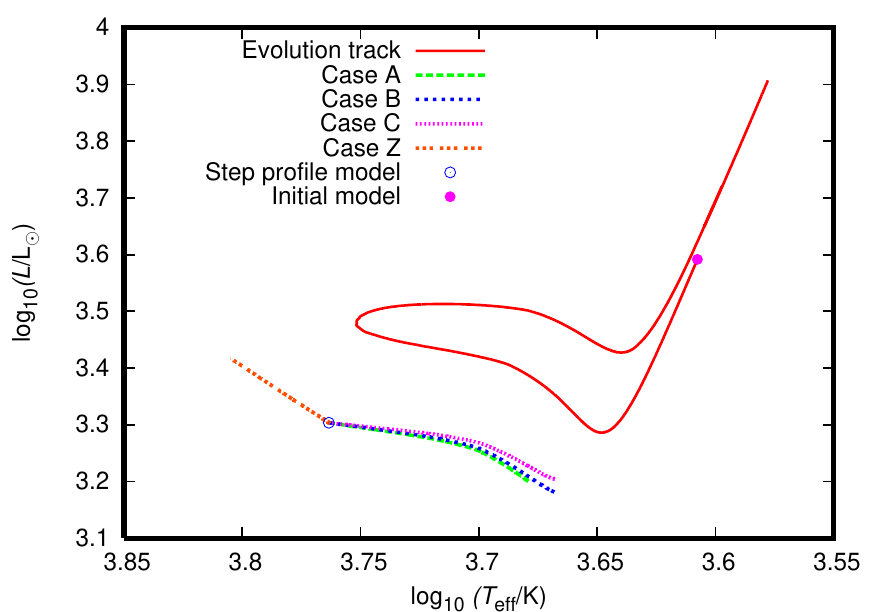}
\caption{A series of models in which half the hydrogen of the step
  profile model is replaced with arbitrarium out to different
  limits. The arbitrarium is initially identical to hydrogen but
  gradually takes on the opacity of helium.}
\label{fig:opaq2}
\end{figure}

The effect of opacity is simply tested. The code uses opacity tables
that depend on the abundance of hydrogen and helium and the
metallicity. For opacity only we set the mass of helium to be

\begin{equation}
M_{\rm He(opacity)} =  M_{\rm He} + (1-P)M_{\rm Arb}
\end{equation}

\noindent
and that of hydrogen to be

\begin{equation}
M_{\rm H(opacity)} =  M_{\rm H} + P M_{\rm Arb},
\end{equation}

\noindent
where Arb refers to arbitrarium and $P$ is a parameter that we vary
between 0 and 1. When $P$ is zero the arbitrarium contributes to the
opacity in the same way as helium. We gradually increase it to one in
small steps, with the stepsize sufficiently small that thermal
equilibrium is rapidly achieved after each change. The conversion of
all chemical species is suspended so the models are otherwise
static. When $P=1$ the arbitrarium contributes to the opacity in
precisely the same way as hydrogen but is identical to helium in every
other respect. The path in the HR diagram is shown in
Fig.~\ref{fig:opaq} together with the track for normal evolution, the
initial $6\,M_{\odot}$ model and a model with a helium core of the
same size but a step profile composition (Fig.~\ref{fig:prof3}). The
step profile model was created by gradually making the arbitrarium
hydrogen-like in every respect.

It is apparent that the increase in the opacity occasioned by
replacing helium with hydrogen causes the star to move somewhat down
the Hayashi line. However it is still very much redder than both the
tip of the blue loop under normal evolution and a step profile model
with a helium core of the same size. Opacity is part of the
explanation for looping but only a minor component. If the conversion
of chemical species is resumed then the final model performs a blue
loop similar to that observed in the normal evolution track. The point
of maximum temperature is at $3.8$.

We explore the consequences of changing the opacity a little further
by making some additional models. We take the step profile model and
replace half the hydrogen with arbitrarium from the innermost edge at
$1.07\,M_{\odot}$ out to four different masses. These are identified
as Case A, $1.25\,M_{\odot}$; Case B, $1.5\,M_{\odot}$; Case C, $2$
$M_{\odot}$ and Case Z, the surface. The arbitrarium is initially
identical to hydrogen but is gradually given the opacity of
helium. The results are shown in Fig.~\ref{fig:opaq2}. Case Z shows
that if one lowers the overall opacity of the envelope one compels the
star to heat up and move leftwards. Changes near to the burning shell
have the opposite effect and Case B is the reddest. This must mark the
approximate point at which the effect of changing the local opacity
reverses. Contrary to the view of \citet{1990ApJ...364..527S}, a
reduction in the opacity above the burning shell does not necessarily
result in an increase in luminosity. In all cases the behaviour is
that of regime III: a reduction in temperature accompanies a reduction
in luminosity.

\subsection{Fuel Supply}

Replacing helium with hydrogen increases the available fuel supply in
the shell. We test the effect of this by returning to the initial
model and modifying the code so that the local mass of arbitrarium is
added to that of hydrogen for the purposes of calculating the nuclear
reaction rates. As before the parameter $P$ is set to zero and slowly
increased to one.

\begin{equation}
M_{\rm H(rates)} =  M_{\rm H} + P M_{\rm Arb}.
\end{equation}
\noindent

\begin{figure}
\centering
\includegraphics[width=0.5\textwidth]{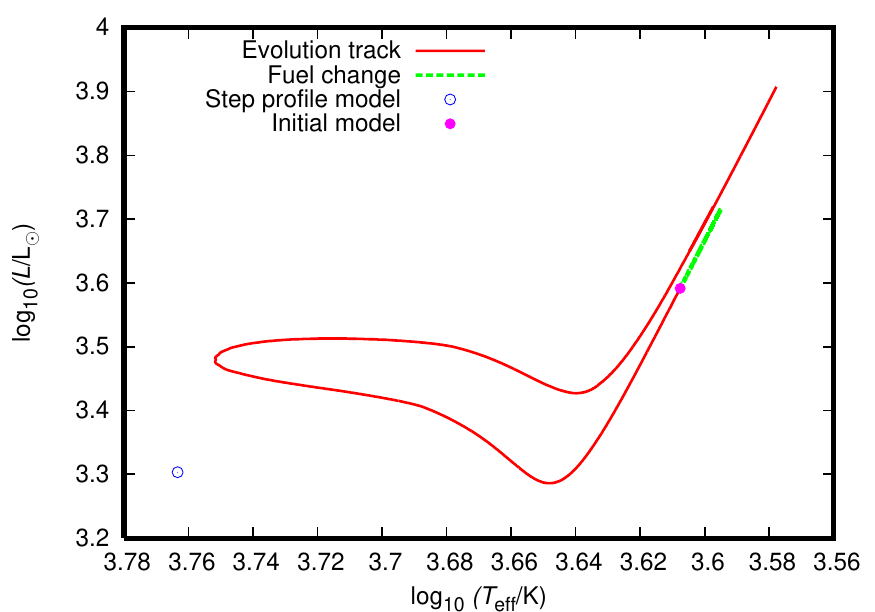}
\caption{The effect of making the arbitrarium count for an equivalent
  mass of hydrogen for burning purposes. It is initially identical to
  helium.}
\label{fig:fuel}
\includegraphics[width=0.5\textwidth]{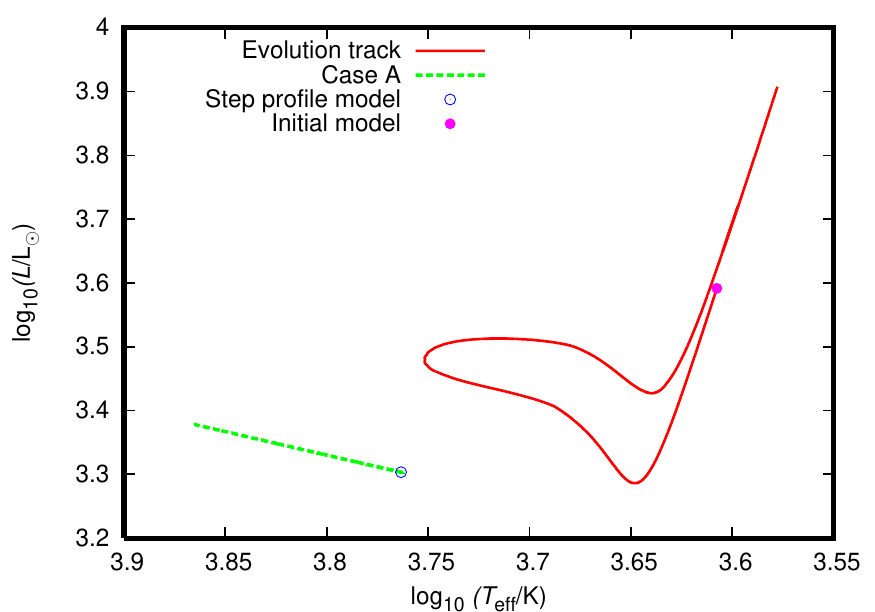}
\caption{Case A is for a model in which half the hydrogen of the step
  profile model is replaced with arbitrarium out to $ 1.25
  M_{\odot}$. The arbitrarium is initially identical to hydrogen but
  its contribution to the hydrogen reaction rates is gradually reduced
  to that of helium, i.e. zero.}
\label{fig:fuel2}
\end{figure}

The effect is shown in Fig.~\ref{fig:fuel}. It is apparent that
increasing the fuel supply propels the star up the Hayashi line and
therefore disfavours looping. This is not too surprising; the shell
homology relations of \citet{1970A&A.....6..426R} have, for stars on
the Hayashi line, the luminosity proportional to a low but positive
power of the hydrogen mass fraction.

As with the opacity we consider the effect of modifying the step
profile model by replacing half the hydrogen with arbitrarium out to
different masses. The situation is simplified by the fact that an
alteration to the arbitrarium only has an effect within the burning
shell. This is very thin, extending from $1.07 M_{\odot}$ out to $1.14
M_{\odot}$. The arbitrarium is initially identical to hydrogen but its
contribution to the reaction rates is gradually reduced to zero. In
Fig.~\ref{fig:fuel2} only Case A is shown, the others are
identical. It is apparent that reducing the fuel supply propels the
star to higher temperatures and luminosities, a further demonstration
that the fuel supply acts to suppress looping. This contradicts the
claim by \citet{2004A&A...418..213X}: an increased fuel supply does
not in fact cause the luminosity increase that takes a star that is
performing a blue loop into the Hertzsprung Gap. A final point is that
a reduction in the energy productivity via either the removal of CNO
elements or an alteration to the reaction rates has the same effect.

\subsection{Mean Molecular Weight}

If, in the course of the evolution of a shell-burning star, changes
to the opacity only slightly favour looping and to the fuel supply
suppress it, then, by elimination, the primary cause must be the mean
molecular weight. This depends on the proton number $Z_{i}$, the
atomic weight $A_{i}$, and the mass fraction $X_{i}$ of the
component species. For an ionised gas the formula is

\begin{equation}
\mu=\left(\sum_i\frac{X_{i}(1+Z_{i})}{A_{i}}\right)^{-1}.
\end{equation}

\noindent
The parameters $A_{\rm Arb}$ and $Z_{\rm Arb}$ are defined in terms of
the parameter $P$, which is slowly increased from zero to one.

\begin{equation}
A_{\rm Arb} =  A_{\rm He} + P(A_{\rm H} -  A_{\rm He})
\end{equation}
\noindent
\begin{equation}
Z_{\rm Arb} =  Z_{\rm He} + P(Z_{\rm H} -  Z_{\rm He})
\end{equation}

\begin{figure}
\centering
\includegraphics[width=0.5\textwidth]{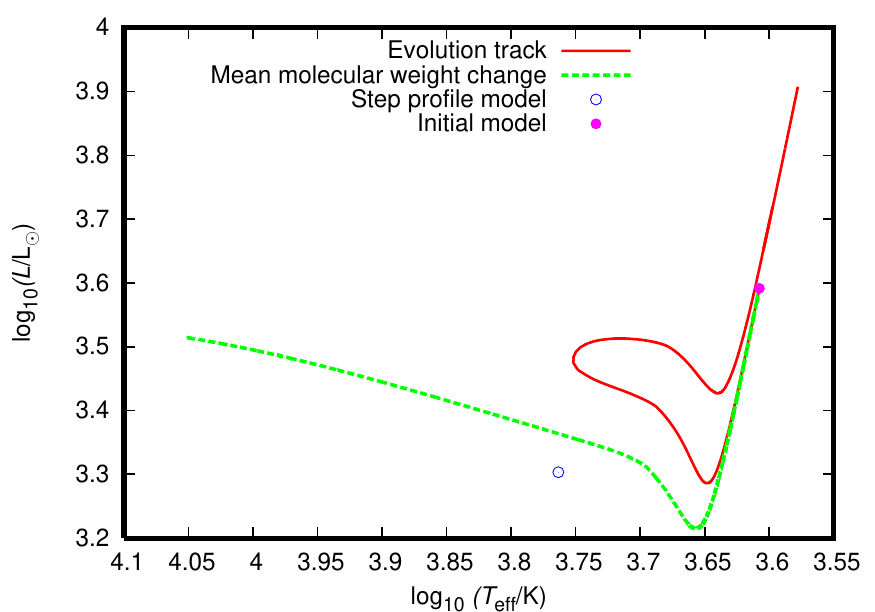}
\caption{The effect of making the arbitrarium count for an equivalent
  mass of hydrogen for the purpose of determining the mean molecular
  weight. The mean molecular weight profile eventually becomes a step
  function.}
\label{fig:mmw}
\includegraphics[width=0.5\textwidth]{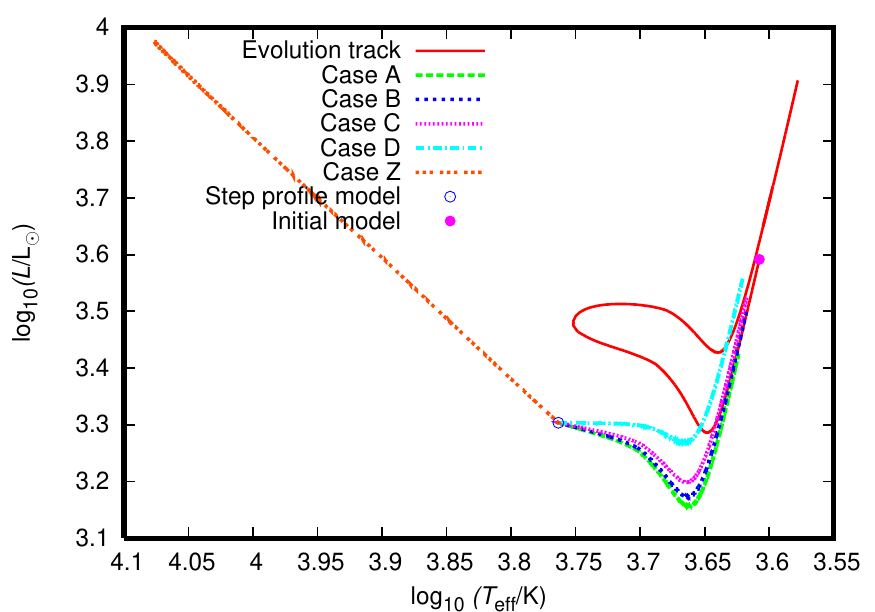}
\caption{A series of models in which half the hydrogen of the step
  profile model is replaced with arbitrarium out to different
  limits. The arbitrarium is initially identical to hydrogen but
  gradually takes on the mean molecular weight of helium. This
  introduces a second step in the mean molecular weight profile.}
\label{fig:mmw2}
\end{figure}

\noindent
The effect is to change the mean molecular weight profile to a step
function and the evolutionary path is shown in Fig.~\ref{fig:mmw}. The
star is propelled down the Hayashi line and far across the Hertzsprung
Gap. The final state is considerably bluer than the step profile
model, confirming our supposition that the mean molecular weight is
the decisive factor.

As before we modified the step profile model by replacing half the
hydrogen with arbitrarium out to different masses. The arbitrarium was
initially identical to hydrogen but gradually assumed the mean
molecular weight contribution of helium. The mean molecular weight of
the helium core was $1.344$ and this fell to $0.642$ in the
envelope. The replacement of half the hydrogen with helium-like
arbitrarium increased the mean molecular weight to $0.868$ and the
results are shown in Fig.~\ref{fig:mmw2}. The labels are the same as
for the opacity with the exception of Case D, for which the
replacement occurred out to $3\,M_{\odot}$. It is apparent that an
increase in the mean molecular weight near the burning shell favours
considerable rightward motion in the HR diagram. The zone of
importance is greater in extent than the burning shell itself but less
than the entire star. Increasing the mean molecular weight throughout
the envelope compels the star to contract and heat up. Finally, if all
other things are equal, an increase in the thickness of the modified
zone causes a monotonic increase in luminosity.

\section{Implications}

Any discussion about the nature of blue loops would be incomplete
without a reference to the observation by \citet{2013sse..book.....K}
that, ``\emph{the present phase is a sort of magnifying glass, also
  revealing relentlessly the faults of calculations of earlier
  phases.}'' It is apparent that, largely through the medium of the
mean molecular weight, anything that affects the composition profile
above the burning shell has a noticeable effect on the extent of the
blue loop. Anything that affects the mass of the core also has an
effect. This is partly because a star whose core is relatively more
massive is redder, partly because a star whose core is absolutely more
massive evolves faster and thus enters the loop-curtailing phase of
rapid contraction faster, perhaps before the shell has burnt through
the composition profile. With this in mind we consider the effect of
changes to the initial helium abundance and metallicity.

\subsection{The Initial  Helium Abundance}

An increase in the initial helium abundance has long been associated
with more extended blue loops
\citep{1971ApJ...170..353R}. \citet{2004A&A...418..225X} attributed
this to the reduction in the overall opacity in the envelope. It is
apparent from Case Z in Fig.~\ref{fig:opaq2} that such a reduction, as
opposed to one localised to the vicinity of the burning shell, does
indeed shift a star bluewards. \citet{2000ApJ...543..955B} cited the
reduced opacity as a cause but also stated that the associated
increase in the mean molecular weight has an effect. From
Fig.~\ref{fig:mmw2} we can see that an overall increase in the mean
molecular weight of the envelope also shifts a star bluewards. The
effect is much more powerful than that exerted by the
opacity. Previous authors have however neglected the final effect of
additional helium in the envelope. More helium means less hydrogen and
therefore a reduction in the fuel supply to the burning
shell. Fig.~\ref{fig:fuel2} shows that this also produces a blueward
shift, one that is greater than that caused by the opacity change but
still less than that caused by changes to the mean molecular weight.

It should be noted that the above analysis pertains to both a fixed
helium core mass and a constant composition profile. To appreciate the
full effect of an increase in the initial helium content we must
consider the evolution of the star from the zero-age main
sequence. The first point to consider is that the helium core produced
by hydrogen burning is more massive. For a $6\,M_{\odot}$ star with
solar metallicity the mass of the helium core at the top of the
Hayashi line increases from $0.94\,M_{\odot}$ when $Y=0.26$ to $0.99$
$M_{\odot}$ when $Y=0.30$. Secondly, the base of the convective region
rises from $1.26\,M_{\odot}$ to $1.36\,M_{\odot}$, meaning that first
dredge-up removes less excess helium.

\begin{figure}
\centering
\includegraphics[width=0.5\textwidth]{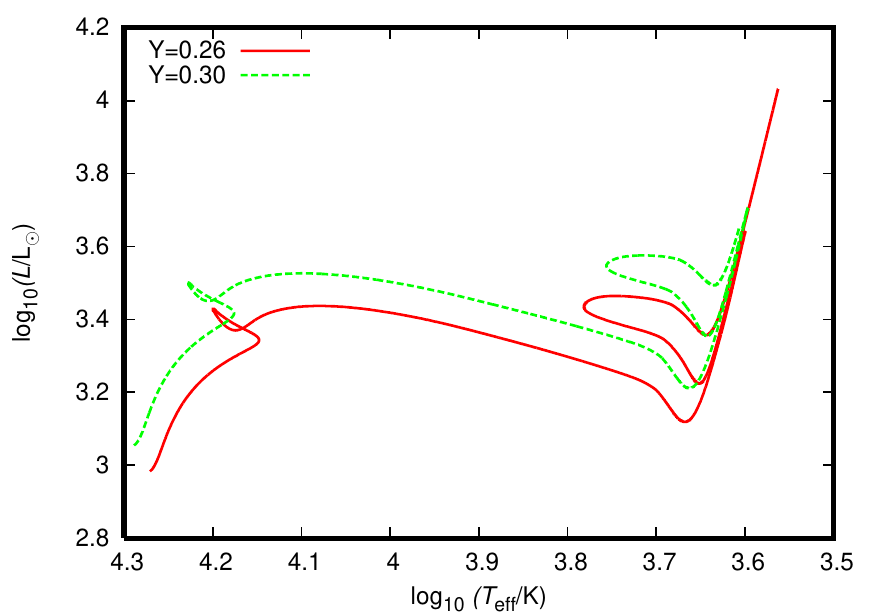}
\caption{Models showing the evolution of two $6 M_{\odot}$ stars with
  $Z=0.02$.}
\label{fig:zhe}
\end{figure}

\begin{figure}
\centering
\includegraphics[width=0.5\textwidth]{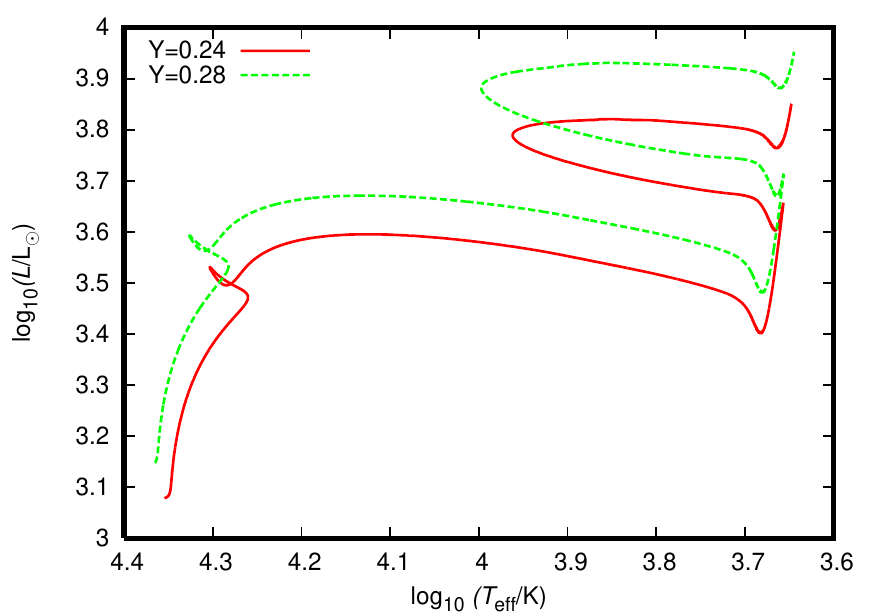}
\caption{Models showing the evolution of two $6 M_{\odot}$ stars with
  $Z=0.002$.}
\label{fig:10zhe}
\end{figure}

If we turn to the stellar models we find that looping, although
enhanced at low metallicity (Fig.~\ref{fig:10zhe}), is in fact
retarded by an increase in $Y$ at high metallicity
(Fig.~\ref{fig:zhe}). This phenomenon was observed by
\citet{2004A&A...418..225X} but they did not attempt to explain
it. The cause is clear: at high metallicity the consequences of a
larger core and less dredging dominate over the envelope effects. At
low metallicity the stars do not move very far up the Hayashi line and
first dredge-up is either limited or non-existent.

\subsection{The Metallicity}

\begin{figure}
\centering
\includegraphics[width=0.5\textwidth]{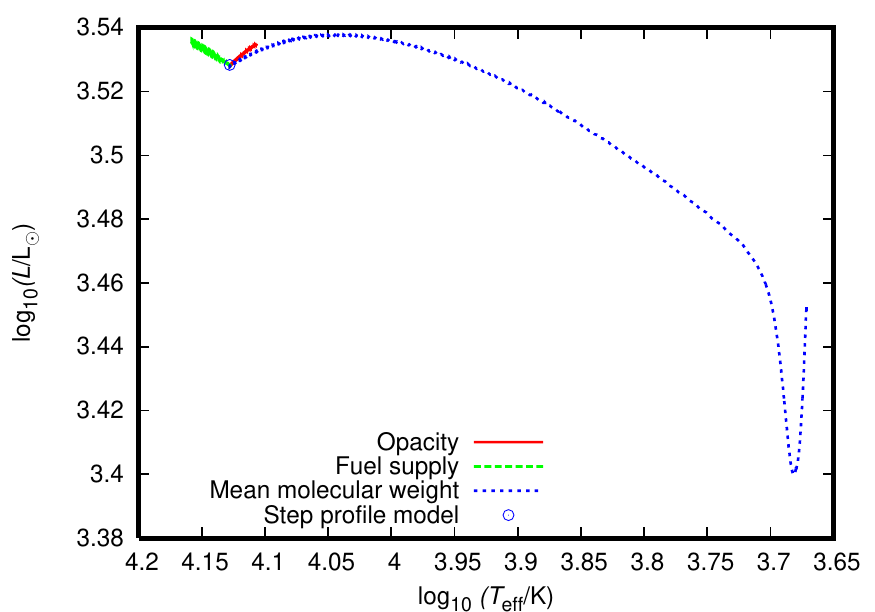}
\caption{Models showing the effect of taking a $6\,M_{\odot}$ step
  function model with $Z=0.002$ and replacing half the hydrogen above
  the shell out to $1.5\,M_{\odot}$ with hydrogen-like
  arbitrarium. This is separately given the opacity, contribution to
  hydrogen-burning (i.e. no contribution) and molecular weight of
  helium}
\label{fig:lowzzz}
\end{figure}

It is sensible to check that the effects of a change to the opacity,
the fuel supply and the mean molecular weight are the same at low
metallicity. To that end we take a 6 $M_{\odot}$ star at tenth-solar
composition ($Z=0.002$ and $X=0.745$) and create a step profile
model. The burning shell is located between $0.95$ and $1.18$
$M_{\odot}$ and we replace half the hydrogen out to $1.5\,M_{\odot}$
with hydrogen-like arbitrarium. This is then separately given the
properties of helium for the three different factors. The results are
shown in Fig.~\ref{fig:lowzzz} and the picture is much the same as
with the solar models.

As with an increase in the initial helium abundance, a reduction in
metallicity has also long been associated with more developed loops
\citep{1971ApJ...170..353R}. \citet{2004A&A...418..225X} performed a
thorough study of the variability of blue loops with metallicity and
helium content and found that, for fixed $Y$ and a mass of
$7\,M_{\odot}$, loops become shorter as $Z$ is increased from $0.004$
to $0.012$. They attributed this to a reduction in the overall opacity
and we decided to test this.

\begin{figure}
\centering
\includegraphics[width=0.5\textwidth]{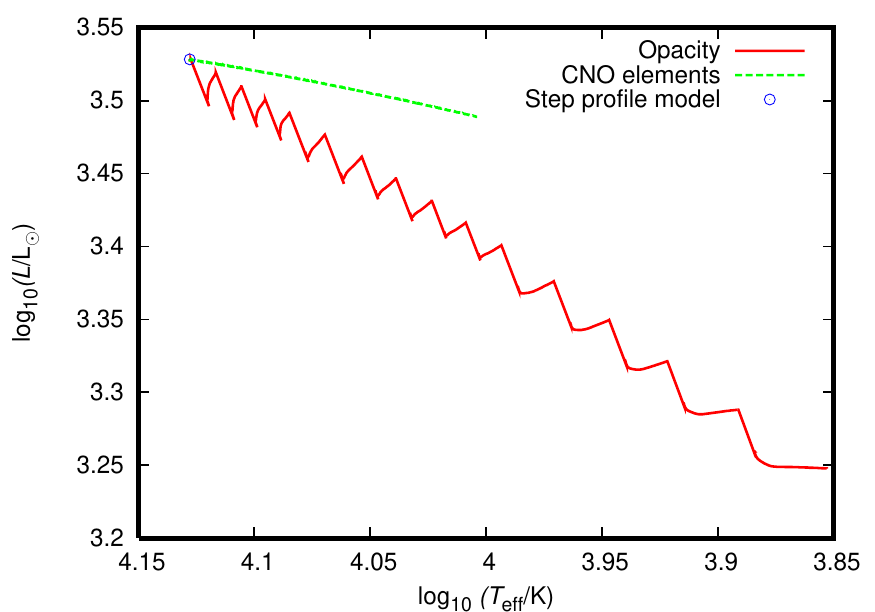}
\caption{Models showing the effect of taking a $6 M_{\odot}$ step
  function model with $Z=0.002$ and giving it either the opacity or
  the CNO cycle enhancement of a star with $Z=0.02$.}
\label{fig:opacityandfuel}
\end{figure}

The ${\sc \rm STARS}$ code uses a different opacity table for each
metallicity and this table is loaded when the program is started. This
means that it is difficult to gradually change the opacity to reflect
different metallicities. We wrote a script to copy in a new opacity
table, start the program, evolve the star to thermal equilibrium,
output the final model as the new input model and repeat. With this
method we can give the $Z=0.002$ step profile model the $Z=0.02$
opacity tables. The sawtooth pattern (Fig.~\ref{fig:opacityandfuel})
marks the sudden drop in luminosity as the table is changed and this
is then followed by the relaxation of the envelope. The intermediate
opacity tables were for $Z=0.0025$, $0.003$, $0.0035$, $0.004$,
$0.005$, $0.006$, $0.007$, $0.008$, $0.009$, $0.01$, $0.012$, $0.014$,
$0.016$, and $0.018$.

It is apparent that the opacity increase associated with an increase
in the metallicity does indeed greatly disfavour looping. Despite this
it is sensible to consider the other consequences of varying the
metallicity. To that end we return to the step profile model and
subtract $0.018$ from the helium mass fraction throughout the
star. This is then replaced with helium-like arbitrarium and, when
added to the metals, gives a mass fraction of $0.02$. We find that
changing the mean molecular weight contribution of the arbitrarium to
that of metals has a negligible effect. This is not very surprising
given that helium and metals have a similar mean molecular
weight. When the arbitrarium does not contribute to helium burning the
effect is much the same. The only modification that does have a
noticeable effect is to allow the arbitrarium to contribute in the
same way as the metals to the CNO cycle. We have already shown that an
increase in the fuel supply in the shell suppresses looping and the
same effect is apparent in Fig.~\ref{fig:opacityandfuel}. The effect
is less important than a change to the opacity but not
inconsequential.

\begin{figure}
\centering
\includegraphics[width=0.5\textwidth]{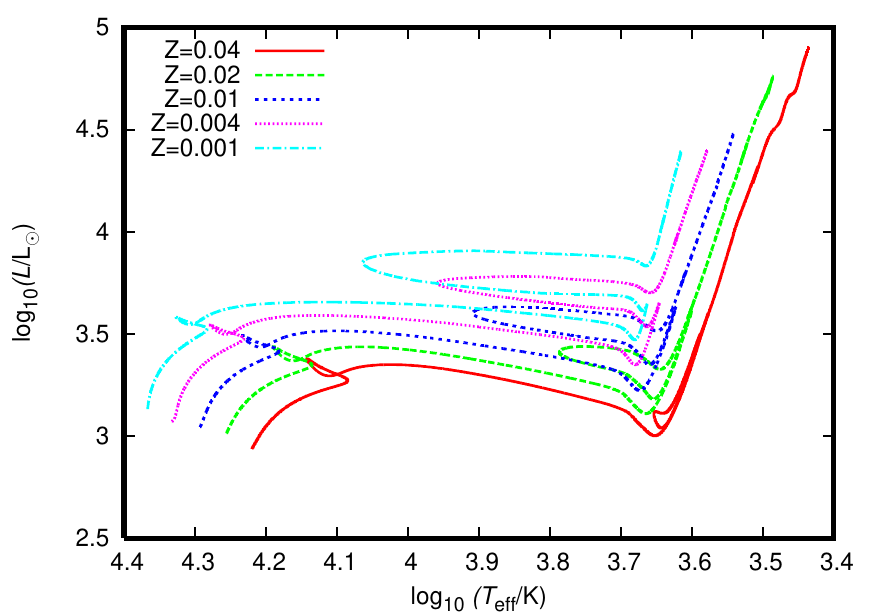}
\caption{Evolution tracks for five $6 M_{\odot}$ stars with $Y=0.25$
  and different metallicities.}
\label{fig:Z6}
\end{figure}

\citet{2004A&A...418..225X} found that if the metallicity is increased
above $Z=0.012$ then their blue loops became longer. They attributed
this to the fact that first-dredge up becomes important for these
stars and that the increased dredging of metal-rich stars favours
looping more than the increased opacity disfavours it. This is
certainly the correct explanation for their findings. However, when we
make a series of $6\,M_{\odot}$ models we find that the extent of
looping decreases monotonically as the metallicity is increased
(Fig.~\ref{fig:Z6}). Evidently different codes cause one to draw
different conclusions. The more recent models of
\citet{2009A&A...508..355B}, \citet{2012A&A...543A.108L} and
\citet{2013A&A...558A.103G} exhibit similar metallicity evolution to
what we observe and therefore the current consensus may be said to be
that blue loops become shorter at higher metallicities.  We will not
go beyond the observation that although the general trend is for
looping to become less pronounced as metallicity is increased, we
would not be surprised to find occasional deviations from this rule.

\section{Conclusions}

We began by considering past work. Previous authors have shown that
excess helium above the burning shell shifts a star to the right in
the HR diagram and makes it redder and larger than it would otherwise
be. The shrinkage of the convective core during hydrogen burning means
that this is the case at the start of helium burning. The outward
movement of the shell in mass reduces the amount of excess helium and
this is accompanied by leftward motion into the Hertzsprung Gap -- the
start of the loop.

Some researchers have been labouring under various misconceptions
about the blue loop mechanism and that these have arisen from flawed
logic. The blue loop is not the result of a thermal instability
triggered by a luminosity drop; rather a star must necessarily feature
a luminosity drop if it is to move back down the Hayashi line. Stars
with very low metallicities can sometimes begin a blue loop in the
Hertzsprung Gap and this is not accompanied by a drop in luminosity.
If a star evolves rapidly then the envelope may become thermally
unstable but this, as with the shrinkage and disappearance of the deep
convective envelope, is an effect rather than a cause.

The presence of excess helium alters the properties of a star in
different ways. Although there has been some speculation, to date
nobody has attempted to make a proper comparison of the different
factors under circumstances that ensure that the effect of their
variation alone is observed. We considered three possibilities: the
opacity, the fuel supply and the mean molecular weight. By modifying
our code to include arbitrarium, an artificial element that possessed
some of the characteristics of hydrogen and some of those of helium,
we are able to compare these factors independently. We find that
changes to the opacity, which is lowered by the excess helium, favours
looping but that the effect is small. An increase in the fuel supply
has the opposite effect and suppresses looping. This leaves the mean
molecular weight, which turns out to be the decisive factor. Changes
to the opacity and mean molecular weight over the whole envelope have
the opposite effect.

We then considered the effect of changes to initial helium abundance
$Y$ and metallicity $Z$ on the extent of looping. We showed that an
increase in $Y$ favours looping at low metallicity but suppresses it
at high metallicity. An increase in $Y$ reduces the overall opacity,
increases the overall mean molecular weight and reduces the fuel
supply in the shell, all of which favour blueward motion in the HR
diagram. However at high metallicity the extent of looping is
strongly affected by the depth of first dredge-up. This is shallower
if the helium content is increased and therefore loops are
shorter.

Similarly an increase in the metallicity increases the opacity and the
efficiency of hydrogen burning and thus suppresses looping. However,
at high metallicity a further increase leads to more extensive
envelope convective zones and deeper dredging. This may succeed in
making the loops bluer. Such complexity and sensitivity probably
accounts for the poor fit of existing models of blue helium-burning
stars to observations \citep{2011ApJ...740...48M}.

\bibliographystyle{mn2e} \bibliography{thesis}

\end{document}